\documentstyle[12pt]{article}
\def\beq{\begin{equation}}
\def\eeq{\end{equation}}
\def\bea{\begin{eqnarray}}
\def\eea{\end{eqnarray}}

\begin {document}
\begin{titlepage}
June 1992 \hfill HU Berlin--IEP--92/3 \\
\mbox{ }  \hfill hepth@xxx/9206053
\vspace{6ex}
\Large
\begin {center}
\bf{On Correlation Functions for Non-critical Strings
with $c\leq 1$ but $d\ge 1$}
\end {center}
\large
\vspace{3ex}
\begin{center}
H. Dorn and H.-J. Otto \footnote{e-mail: dorn@ifh.de or otto@ifh.de}
\end{center}
\normalsize
\it
\vspace{3ex}
\begin{center}
Fachbereich Physik der Humboldt--Universit\"at \\
Institut f\"ur Elementarteilchenphysik \\
Invalidenstra\ss e 110, D--O--1040 Berlin, Germany
\end{center}
\vspace{6 ex }
\rm
\begin{center}
\bf{Abstract}
\end{center}
We construct a Goulian-Li-type continuation in the number of insertions
of the cosmological constant operator which is no longer restricted to one
dimensional target space. The method is applied to the calculation of the
three-point and a special four-point correlation function. Various aspects
of the emerging analytical structure are discussed.
\end {titlepage}
\newpage
\setcounter{page}{1}
\pagestyle{plain}
\section {Introduction}
Till now non-critical string theories have been successfully constructed
for target space dimensions $d\leq 1 $ only. Besides the lattice approach
along conventional lines (see e.g. \cite{c1}) a considerable amount of
attention has
been devoted to conformal matter theories with central
charge $c\leq 1$ coupled to gravity in the formulation as matrix models
\cite{c2}. Attempts to go beyond the $c=1$ barrier in this scheme have
been initiated recently \cite{c3}.

On the other hand following refs. \cite{c4,c5,c6,c7} one can, at least
for the lowest genera, reproduce the matrix model results for correlation
functions in a continuum formulation by manipulations with the defining
functional integrals. The central point in this procedure is a continuation
in the number of insertions of the so called cosmological or sreening
operator. The technique is connected with an extensive use of kinematical
peculiarities in $d=1$ target space dimension.
Although we cannot circumvent the restriction $c\leq 1$, we will be able to
avoid the dependence on the special $d=1$ kinematics. The present letter
gives a short presentation of our continuation procedure for $d\geq 1$,
the proof of equivalence with refs.\cite{c4,c5,c6,c7} for $d=1$ and
some discussion of the emerging analytical structure of our amplitudes.
\section {Formulation of the model}
The action of the combined matter and Liouville system is
\begin{eqnarray}
S&=&\frac{1}{8\pi}\int d^{2}z\sqrt{g(z)}\Big(g^{mn}(z)\partial_{m}X^{\mu}
(z)\partial_{n}X_{\mu}+iR(z)P^{\mu}X_{\mu}(z)\Big)\nonumber\\
&+&\frac{1}{8\pi}\int d^{2}z\sqrt{g(z)}\Big(g^{mn}(z)\partial_{m}\Phi(z)
\partial_{n}\Phi+Q R(z)\Phi(z)+\mu^{2}e^{\alpha\Phi(z)}\Big).
\label{e1}
\end{eqnarray}
Then the central charge $c$ of the matter theory is given by
\begin{equation}
c=d-3P^{2}
\label{e2}
\end{equation}
for a $d$-dimensional target space $(\mu=1,...,d)$.
\footnote{We assume throughout this paper $d<25$}
$Q$ and $\alpha $ are fixed by the requirement
of Weyl invariance of the quantum field theory described by (\ref{e1})
and by the locality requirement for $exp(\alpha\Phi)$ \cite{c8,c9,c10}
\begin{equation}
Q^{2}=\frac{25-c}{3}=\frac{25-d}{3}+P^{2}~~,
\label{e3}
\end{equation}
\begin{equation}
\alpha_{\pm}=\frac{Q}{2}\pm \frac{\sqrt{Q^{2}-8}}{2},~~~  \alpha\equiv
\alpha_{-}~~.
\label{e4}
\end{equation}
We are interested in correlation functions of gravitational dressed local
vertex operators \cite{c8,c9,c11}
\begin{equation}
V_{k} = e^{ikX(z)}e^{\beta\Phi(z)}
\label{e5}
\end{equation}
with
\footnote{If compared to \cite{c11} we use a slightly changed normalization :
$Q^{2}+P^{2}\rightarrow Q^{2},\beta =\frac{2b}{Q},\alpha =\frac{2a_{0}}{Q}$}
\begin{equation}
\beta = \frac{Q}{2} - \sqrt{m^{2}+(k-P/2)^{2}}
\label{e6}
\end{equation}
where
\begin{equation}
m^{2} \equiv \frac{1-d}{12}~~.
\label{meyer0}
\end{equation}
Real $\alpha $ requires $c\leq 1$. This can be realized in target space
dimensions $d>1~~(m^{2}<0)$ for large enough $P^{2}$. $\beta $ is real for
most of the $k$, but there is an exceptional compact domain $E$ whose
boundary is given by $\beta =Q/2$ and which is characterized by
\bea
m^{2}+k_{\perp}^{2}&<&0,
\nonumber\\
-\sqrt{-m^{2}-k_{\perp}^{2}}&<&k_{\parallel}-\frac{\vert P \vert }{2}<
+\sqrt{-m^{2}-k_{\perp}^{2}}
\label{e7}
\eea
with $k_{\parallel},k_{\perp}$ denoting the components of $d$-dimensional
momentum $k$ parallel and orthogonal to $P$. We start out with $k\notin E$
and continue afterwards.
\section{Three point function}
The three point function has been calculated in \cite{c5,c7} for positive
integer $s$
\beq
\alpha s =Q-\sum_{j=1}^{3}\beta_{j}
\label{e8}
\eeq
\bea
A_{3}=\delta (\sum_{j}k_{j}-P)\frac{\Gamma (-s)}{\alpha}\Gamma (1+s)
\Big(\frac{\mu ^{2}}{8}\frac{\Gamma (1+\frac{\alpha ^{2}}{2})}{
\Gamma (-\frac{\alpha ^{2}}{2})}
\Big)^{s}~~\times ~~~~~~~~~~~~~~~~~~~~~~~~~~~~~~~~~~~~~~~\nonumber\\
exp\Big \{f(-\frac{\alpha ^{2}}{2},-\frac{\alpha ^{2}}{2}\vert s)-
f(1+\frac{\alpha ^{2}}{2},\frac{\alpha ^{2}}{2}\vert s)
+\sum_{j=1}^{3}\Big(
f(1-\alpha \beta_{j},-\frac{\alpha ^{2}}{2}\vert s)-
f(\alpha \beta_{j},\frac{\alpha ^{2}}{2}\vert s)\Big)\Big\}.
\label{e9}
\eea
The infinity due to the pole in $\Gamma(-s)$ is understood as a signal of the
logarithmic modification of the $\mu $ -scaling law for positive integer $s$
\cite{c7,c11}. We introduced $f(a,b\vert s)$ as
\beq
f(a,b\vert s)=\sum_{j=0}^{s-1}\log \Gamma (a+bj).
\label{e10}
\eeq
This function fulfills the relations
\beq
f(a,b\vert s+1)=f(a,b\vert s)+\log \Gamma (a+bs)~~,
\label{e11}
\eeq
\beq
f(a+1,b\vert s)=f(a,b\vert s)+s\log b +\log \Gamma (\frac{a}{b}+s)-
\log \Gamma (\frac{a}{b})~~,
\label{e12}
\eeq
\beq
f(a+b,b\vert s)=f(a,b\vert s)+\log \Gamma (a+bs)-\log \Gamma (a)~~,
\label{e13}
\eeq
\beq
f(a+b(s-1),-b\vert s)=f(a,b\vert s)~~,
\label{e14}
\eeq
\beq
f(2a,2b\vert s)=f(a,b\vert s)+f(a+\frac{1}{2},b\vert s)+
(s(s-1)b+s(2a-1))\log 2 -\frac{s}{2}\log \pi~~,
\label{e15}
\eeq
\beq
f(a,0\vert s)=s\log \Gamma (a)~~,
\label{e16}
\eeq
\beq
f(a,\frac{1}{s}\vert s)=\frac{1}{2}(s-1)\log 2\pi+(\frac{1}{2}-sa)\log s
+\log \Gamma(sa).
\label{e17}
\eeq
We have found a continuation of $f$ to arbitrary $s$, which is given by the
following integral representation
\bea
f(a,b\vert s)=\int_{0}^{\infty }\frac{dt}{t}\Big(s(a-1)
e^{-t}+b\frac{s(s-1)}{2}
e^{-t}-s\frac{e^{-t}}{1-e^{-t}}
\nonumber\\
+\frac{(1-e^{-tbs})e^{-at}}{(1-e^{-tb})
(1-e^{-t})}\Big).
\label{e18}
\eea
By careful manipulations with this integral representation one can show, that
our continuation fulfills all the functional relations (\ref{e11}-\ref{e17})
also for generic complex $a,b,s$.
The integral representation (\ref{e18}) is convergent for positive real parts
of $a,b,s$. Together with the functional relations this fixes the complete
analytical structure of $f(a,b\vert s)$. Due to (\ref{e14}) we can restrict
ourselves to $Re b>0$ . Then we find via (\ref{e11}) and (\ref{e12}) that
$exp(f(a,b\vert s)$ has poles at
\beq
a=-bj-l~~~~~~~(poles)
\label{e19}
\eeq
and zeros at
\beq
a+bs=-bj-l~~~~(zeros)~~,
\label{e20}
\eeq
with $Re ~b>0;~~j,l=0,1,2,...~.$
The order of poles and zeros is determined by the number of different
realizations of the r.h.s. of eqs. (\ref{e19}) and (\ref{e20}),
respectively.
Using (\ref{e14}) we get from (\ref{e9})
\beq
A_{3}=\delta (\sum_{j}k_{j}-P)\frac{\Gamma (-s)}{\alpha}\Gamma (1+s)
\Big(\frac{\mu ^{2}}{8}\frac{\Gamma (1+\frac{\alpha ^{2}}{2})}{
\Gamma (-\frac{\alpha ^{2}}{2})}
\Big)^{s}
\prod_{i=0}^{3}F_{i}~~,
\label{e21}
\eeq
where we introduced
\bea
\overline {\beta_{i}}\equiv 1/2(\beta _{j}+\beta _{k}-\beta _{i})=
1/2(Q-\alpha s)-\beta _{i}~~,
\nonumber\\
(i,j,k)~=~perm(1,2,3)~~,
\label{meyer1}
\eea
\beq
F_{i}~=~exp\{ f(\alpha \overline{\beta _{i}},\frac{\alpha ^{2}}{2}\vert s)
-f(\alpha \beta _{i},\frac{\alpha ^{2}}{2}\vert s)\}
\label{e22}
\eeq
and
\beq
\alpha \beta _{0}~=~1+\frac{\alpha ^{2}}{2},~~~~~\alpha \overline{\beta _{0}
}~=~-\frac{\alpha ^{2}s}{2}.
\label{e23}
\eeq
Eqs. (\ref{e19}) and (\ref{e20}) determine the poles and zeros of $F_{i}$
as follows
\beq
2\beta _{i}~=~\alpha _{+}l_{i}~+~\alpha _{-}(j_{i}-s)~~~(poles),
\label{e24}
\eeq
\beq
2\beta _{i}~=~\alpha _{+}l_{i}~+~\alpha _{-}j_{i}~~~~~~~(zeros).
\label{e25}
\eeq
The integers $j_{i},l_{i}$ have to be both $\leq 0$ or both $>0$, i.e.
\beq
(j_{i}~-~1/2)~(l_{i}~-~1/2)~>~0.
\label{e26}
\eeq
For general $s$ poles and zeros are located at different values of
$\beta _{i}$, but if $\alpha _{-}s~=~\alpha _{-}m~+~\alpha _{+}n$ with integer
$m,n$
a lot of poles and zeros cancel.

Of special interest for our later discussion is integer $s>0$. In this this
case the remaining pole-zero pattern is given by
\beq
2\beta _{i}~=~\alpha _{+}(l_{i}+1)~-~\alpha _{-}j_{i}~~~(poles)~~,
\label{e27}
\eeq
\beq
2\beta _{i}~=~-\alpha _{+}l_{i}~-~\alpha _{-}j_{i}~~~~~~(zeros)~~,
\label{e28}
\eeq
with
\beq
integer~s~\geq 0;~~~~j_{i}~=~0,1,...,s-1;~~~~l_{i}~=~0,1,...\infty.
\label{e29}
\eeq
In contrast to the general case the pole positions in $\beta _{i}$
are bounded from below.

Such boundedness is achieved also for $s>0$ with
\beq
\alpha _{-}s~=~\alpha _{+}n~+~\alpha _{-}([s]-m);~~~~1~\leq ~n~\leq ~m~
\leq ~[s].
\label{meyer2}
\eeq

Special care is needed for $F_{0}$. Due to (\ref{e23}) for all $s$ we
are just sitting on the $j_{0}=l_{0}=1$ zero of (\ref{e25}). On the other
hand, for positive integer $s$, where we start our continuation, due to
pole-zero cancellation $F_{0}$ is finite. A finite $F_{0}$ emerges also in
the more general situation
\beq
\alpha _{-}s~=~\alpha _{+}(l_{0}-1)~+~\alpha _{-}(j_{0}-1)~~,
\label{meyer3}
\eeq
with $l_{0},j_{0}$ fulfilling (\ref{e26}).

In all cases where pole-zero cancellation does not guarantee finite $F_{0}$
we treat it using (\ref{e12}) as
\beq
F_{0}~=~-(\frac{\alpha ^{2}}{2})^{-s}\frac{\pi}{\Gamma (1+s)}exp~\Big[ f(1
-\frac{\alpha ^{2}}{2}s,\frac{\alpha ^{2}}{2}\vert s)-f(1+\frac{\alpha
^{2}}{2},
\frac{\alpha ^{2}}{2}\vert s)\Big] ,
\label{e30}
\eeq
where a factor $\Gamma (0) sin~\pi s$ has been set equal to 1.
Of course, this procedure still has to be justified by comparing the $s$-
asymptotics of the resulting expression with that of the original functional
integral. (For $d=1$ see ref \cite{c12}.)

The analytic structure af $A_{3}$ beyond the trivial $s$-dependence in $F_{0}$
and the first factors in (\ref{e21}) is determined by that of $\prod_{i=1}
^{3}F_{i}$. Eliminating $s$ in (\ref{e24}) by using (\ref{meyer1}) we get
finally
\beq
2\overline{\beta _{i}}=-\alpha _{-}(j_{i}-1)-\alpha _{+}(l_{i}-1)~~~(poles)~~,
\label{e31}
\eeq
\beq
2\beta _{i}=\alpha _{-}j_{i}+\alpha _{+}l_{i}~~~~~~~~~~~~(zeros)~~,
\label{e32}
\eeq
with $j_{i},l_{i}$ respecting (\ref{e26}).

In contrast to the $d=1$ case, which we are going to discuss in the next
section, for general $d>1$ the pole structure does not factorize into leg
poles. An attempt to interprete the pole structure will be made in section 5.
\section{The case $d=1$}
For $d>1$ the three $\beta _{i}$'s are independent. For $d=1$ there remains no
angular degree of freedom in $k$-space. The defining equation (\ref{e6})
reduces to
\beq
\beta _{i}=\frac{Q}{2}-(k_{i}-\frac{P}{2})\epsilon _{i}~;~~~~
\epsilon _{i}=sign(k_{i}-\frac{P}{2})~~.
\label{e33}
\eeq
Then momentum conservation $\sum k_{i}=P$ induces constraints among the
$\beta _{i}$'s \cite{c7}.
The $\epsilon _{i}$ must be either all equal to one another but opposite
to $sign(P)$ or of the type $(++-)$ or $(--+)$: One finds in the first case
for $i$=1,2,3
\beq
2\overline{\beta _{i}}=-2\beta _{i}+\frac{3Q+P}{2}~~.
\label{e34}
\eeq
For $(++-)$ one gets instead
\beq
2\overline{\beta _{1}}=2\beta _{2}-\frac{Q+P}{2},~~~
2\overline{\beta _{2}}=2\beta _{1}-\frac{Q+P}{2},~~~
2\overline{\beta _{3}}=\frac{Q+P}{2}~~.
\label{e35}
\eeq
Finally, for the case $(--+)$ one has to replace in (\ref{e35}) $P$ by $-P$.

In all cases the position of any pole or zero of $\prod_{i=1}^{3}F_{i}$
becomes a function of a single $\beta_{i}$. To be more explicit, we
consider the situation studied in detail in \cite{c7}, i.e. $(++-)$ and $P<0$.
Then $2\alpha =Q+P$ and using eqs.(\ref{e12}),(\ref{e13}),(\ref{e22})
and (\ref{e30}) we get from
(\ref{e21})
\beq
A_{3}=-\pi \delta (\sum k_{i}-P)\frac{1}{\alpha }
\Big(-\frac{\alpha ^{2}\mu ^{2}}{16}\Delta (\frac{\alpha ^{2}}{2})\Big)^{s}
\prod_{i=1}^{3}\Delta(m_{i})~~,
\label{e36}
\eeq
with $\Delta (x)=\frac{\Gamma (x)}{\Gamma (1-x)},~~
m_{i}=(\beta _{i}^{2}-k_{i}^{2})/2$. This coincides with \cite{c7} up to a
trivial factor $-\pi (-\frac{\alpha ^{2}}{2})^{s}$.
\section{Four point function}
The calculation of the general 4-point function for $d>1$ requires integrals
not available in \cite{c7,c13}. Therefore, as a first step, we consider
\beq
A_{4}(k_{1},k_{2},k_{3},k_{4}=0)=\frac{d}{d\mu ^{2}}A_{3}(k_{1},k_{2},k_{3})=
\frac{s}{\mu ^{2}}A_{3}(k_{1},k_{2},k_{3}).
\label{e37}
\eeq
This seems to be still sufficient to explore the spectrum of the theory by
analyzing the singularities e.g. in the (1,2)-channel. Furthermore, since
(\ref{e6}) together with the momentum conservation implies for any permutation
$(i,j,l)$ of (1,2,3)
\beq
2k_{i}k_{j}=
(\beta _{l}-\frac{Q}{2})^{2}-(\beta _{i}-\frac{Q}{2})^{2}-
(\beta _{j}-\frac{Q}{2})^{2}+m^{2}+\frac{P^{2}}{4}
\label{e38}
\eeq
any candidates for Mandelstam variables of the special 4-point function
(\ref{e37}) can be expressed in terms of $\beta _{1},\beta _{2},\beta _{3}$.
If one defines with a $d$-dimensional picture in mind
\beq
t_{12}=(k_{1}+k_{2}-\frac{P}{2})^{2}=(k_{3}-\frac{P}{2})^{2}=t_{34}
\label{e39}
\eeq
etc., the pole positions (\ref{e31}) are determined by
\beq
\sqrt{t_{23}+m^{2}}-\sqrt{t_{13}+m^{2}}-\sqrt{t_{12}+m^{2}}=
\alpha _{-}(j-\frac{1}{2})+\alpha _{+}(l-\frac{1}{2}).
\label{e40}
\eeq
To get a more familiar picture we now use the correspondence
to $(d+1)$-dimensional critical strings \cite{c7}. For $\mu^{2} =0$
the amplitudes are equal to that of a $(d+1)$-dimensional critical string
theory with a conservation law
$$
\sum k_{i}=P,~~~~\sum \beta_{i}=Q
$$
and eq. (\ref{e6}) can be interpreted as a mass shell condition
\beq
(\beta -\frac{Q}{2})^{2}-(k-\frac{P}{2})^{2}=m^{2}~~.
\label{e41}
\eeq
The corresponding Mandelstam variables are
$$
T_{ij}^{0}=(\beta _{i}+\beta _{j}-\frac{Q}{2})^{2}-
(k_{i}+k_{j}-\frac{P}{2})^{2}~~.
$$
Also for $\mu ^{2} \neq 0$ but integer $s_{4}\geq 0$, with
\beq
\alpha s_{4}=Q-\sum_{i=1}^{4}\beta _{i}~~,
\label{e42}
\eeq
a $(d+1)$-dimensional interpretation is possible. The 4-point function then
turns into a $(4+s_{4})$-point function with $k_{5}=k_{6}=...=k_{4+s_{4}}=0$.
Motivated by these special cases
we define in general
\beq
T_{ij}=(\beta _{i}+\beta _{j}-\frac{Q-\alpha s_{4}}{2})^{2}-
(k_{i}+k_{j}-\frac{P}{2})^{2}.
\label{e43}
\eeq
This ensures $T_{12}=T_{34}$ etc. and fulfills
\beq
T_{12}+T_{13}+T_{14}=4m^{2}-\frac{1}{4}(Q^{2}-P^{2}+\alpha s_{4}
(\alpha s_{4}+2Q)).
\label{e44}
\eeq
Applying this general definition to our special case (\ref{e37}) we get
$(k_{4}=0, \beta _{4}=\alpha, s_{4}=s-1)$
\beq
T_{i4}~=~m^{2}-\alpha (1+\frac{s_{4}}{2})(2\overline{\beta _{i}}+
\frac{\alpha s_{4}}{2})~;~~~~~i=1,2,3.
\label{e45}
\eeq
This gives a transcription of the pole structure (\ref{e31}) of the
3-point function into poles in the Mandelstam variables (\ref{e43})
of the (special) 4-point function. At least under the constraint of
fixed $ T_{14}+T_{24}+T_{34} $, i.e. fixed $ s_{4}=s-1 $ the poles (\ref
{e31}) turn directly into poles in the individual Mandelstam variables.
The resulting spectrum is for general $s$ unbounded in both directions.

For the particular case of fixed integer $s\geq 0$, due to the pole-zero
cancellation leading to (\ref{e27}),(\ref{e28}) we find poles in e.g.
$T_{12}=T_{34}$ only for
\beq
T_{34}~=~m^{2}~+~(2+s_{4})\Big(l-\frac{\alpha ^{2}}{2}(j-\frac{s_{4}}{2})\Big).
\label{e46}
\eeq
These poles can be seen to be due to the standard poles at $m^{2}+2l,
{}~integer~ l\geq 0$ in those channels of the $(4+s_{4})$-point amplitude
which are
constituted by the momenta $k_{1}, k_{2}$ and $j$ times $k=0$.
The pole spectrum in $T_{34}$ is bounded from below also for values of
$s$ fulfilling (\ref{meyer2}). It is remarkable that than in addition
$F_{0}$ is well defined without any modification.

\section{Conclusions}
Since the continuation from a discrete set of points is mathematically
ambiguous our procedure still has to be proven to yield the correct
solution for the physical problem under discussion. This can be done
by comparing the $\vert s \vert \rightarrow \infty $ asymptotics with
that of the original functional integral. The corresponding problem
in one dimensional target space has been solved in ref.\cite{c12}.
Unfortunately the technique of that paper cannot directly be applied
here as it depends essentially on the discreteness of the momentum
spectrum in minimal models.

Nevertheless we are confident to have presented a correct continuation
procedure. First for $d=1$ it reduces to that of refs. \cite{c5}, \cite{c7},
\cite{c12} and second, the resulting analytical structure is already
completely determined by the functional equations for
$f$ and the absence of singularities for $Re~(a,b,s)\geq 0$.

The interpretation of the spectrum is still not completely clear. While
the pole structure for integer $s\geq0$ fits into the $(d+1)$-dimensional
critical string picture well known from the case of vanishing Liouville
mass $\mu$, for general $s$ the pole-zero cancellation enabling just
this interpretation is removed and an unacceptable spectrum emerges.
This may either signal a hidden disease of our model or only a breakdown
of its $(d+1)$-dimensional interpretation. Favouring the last alternative
one has to analyse eq. (\ref{e40}) and to think about mass shell conditions
in $d$-dimensional space induced by quantization of the $\beta _{i}$
values due to the Liouville dynamics.

A last comment concerns the values of $s$ respecting (\ref{meyer2}) at
which the spectrum is at least bounded from below. These noninteger
values of $s$ belong to the set which can be screened by
insertions of screening operators constructed with $\alpha_{-}$ and
$\alpha_{+}$ \cite{c6}.

\vspace*{1cm}
\noindent
{\bf Acknowledgements:}

We thank J. Distler, V.S. Dotsenko, L. Palla, J. Schnittger and G. Weigt
for useful discussions.

\end{document}